\documentclass[entropy,article,accept,moreauthors,12pt,a4paper]{mdpi}
\lastpage{x}
\doinum{10.3390/------}
\pubvolume{xx}
\pubyear{2017}
\history{Received: xx / Accepted: xx / Published: xx}

\Title{Oscillations in multiparticle production processes}

\Author{Grzegorz Wilk$^{1,}$*, Zbigniew W\l odarczyk$^{2}$}

\address{%
$^{1}$ National Centre for Nuclear Research,~Department of Fundamental Research, Ho\.za 69, 00-681; Warsaw, Poland\\
$^{2}$ Institute of Physics, Jan Kochanowski University, \'Swi\c{e}tokrzyska 15; 25-406 Kielce, Poland}

\corres{grzegorz.wilk@ncbj.gov.pl, +48-22-621 60 85}

\abstract{
We discuss two examples of oscillations apparently hidden in some experimental results for high energy multiparticle production processes: (i) - the log-periodic oscillatory pattern decorating the power-like Tsallis distributions of transverse momenta, (ii) - the oscillations of the modified combinants obtained from the measured multiplicity distributions. Our calculations are confronted with $pp$ data from the Large Hadron Collider. We show that in both cases these phenomena can provide new insight into the dynamics of these processes.
}


\begin{document}

\section{Introduction}
\label{sec:I}

In this work we argue that closer scrutiny of the available experimental results from Large Hadron Collider (LHC) experiments can result in some new, so far unnoticed (or underrated), features, which can provide new insight into the dynamics of the processes under consideration. More specifically, we shall concentrate on multiparticle production processes at high energies and on two examples of hidden oscillations apparently visible there: $(i)$ - the log-periodic oscillations in data on the large transverse momenta spectra, $f\left(p_T\right)$ (presented in Section \ref{sec:LPO}) and $(ii)$ - the oscillations of some coefficients in the recurrence relation defining the multiplicity distributions, P(N) (presented in Section 3).

\begin{figure}[h]
\begin{center}
\includegraphics[scale=0.333]{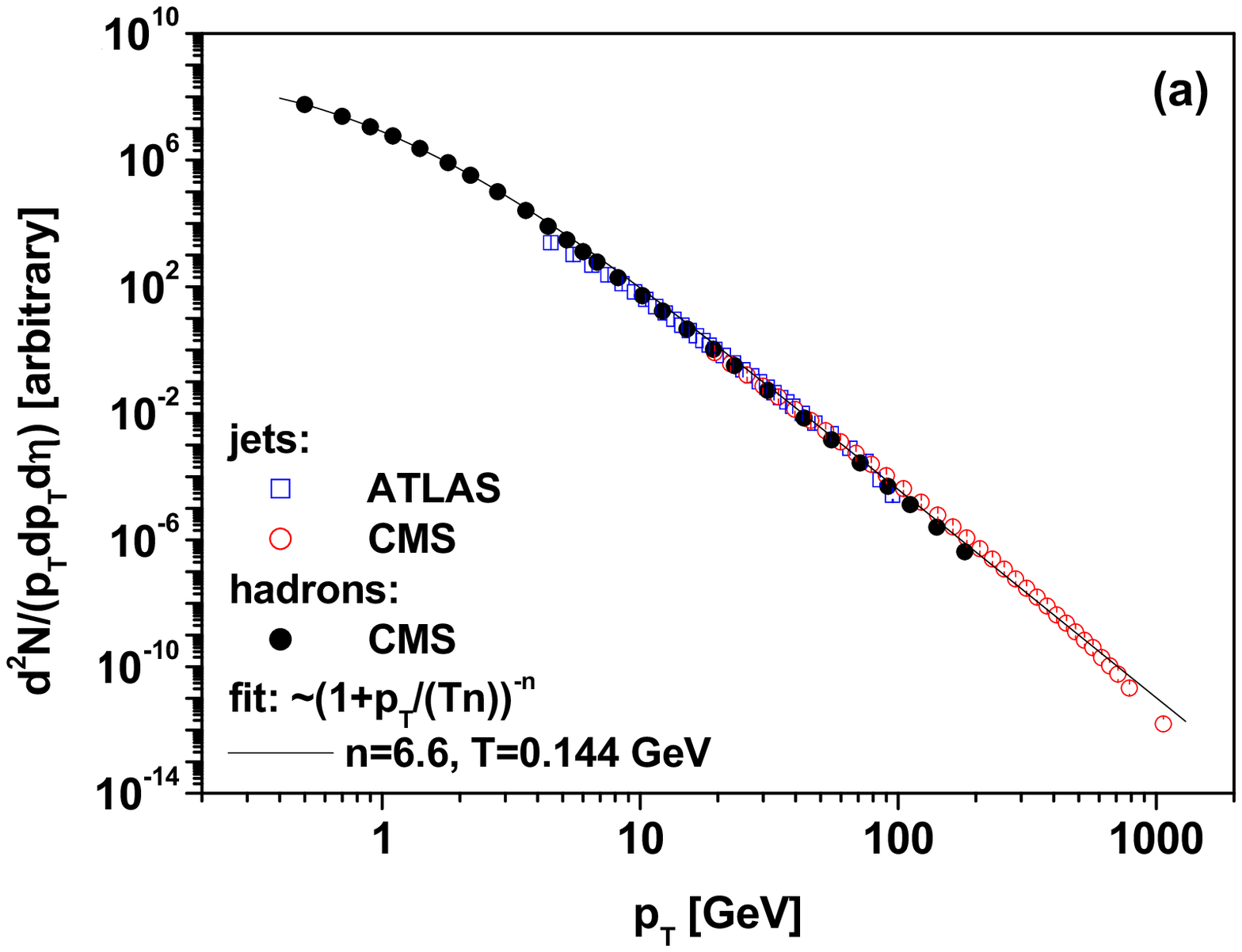}
\includegraphics[scale=0.333]{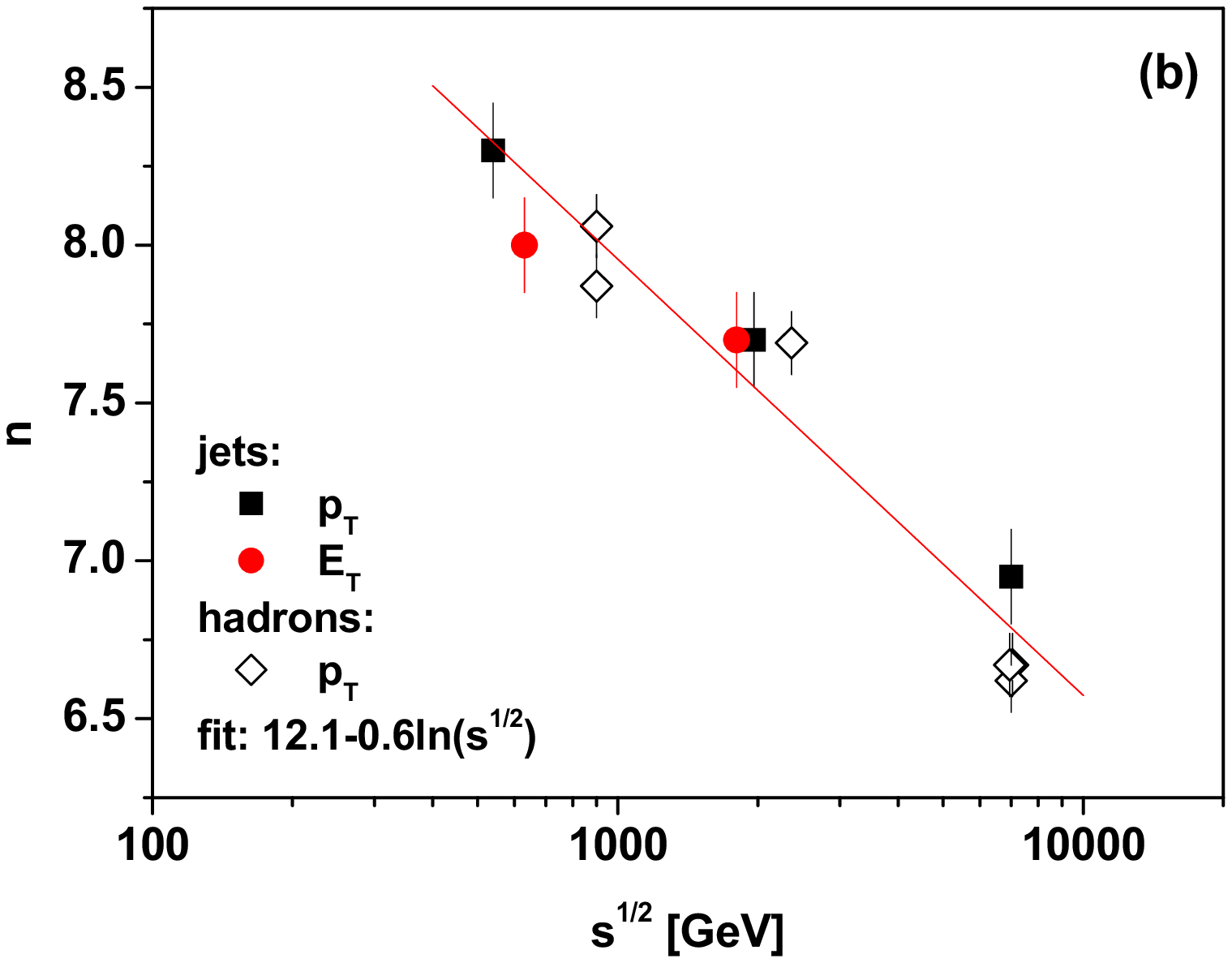}
\includegraphics[scale=0.333]{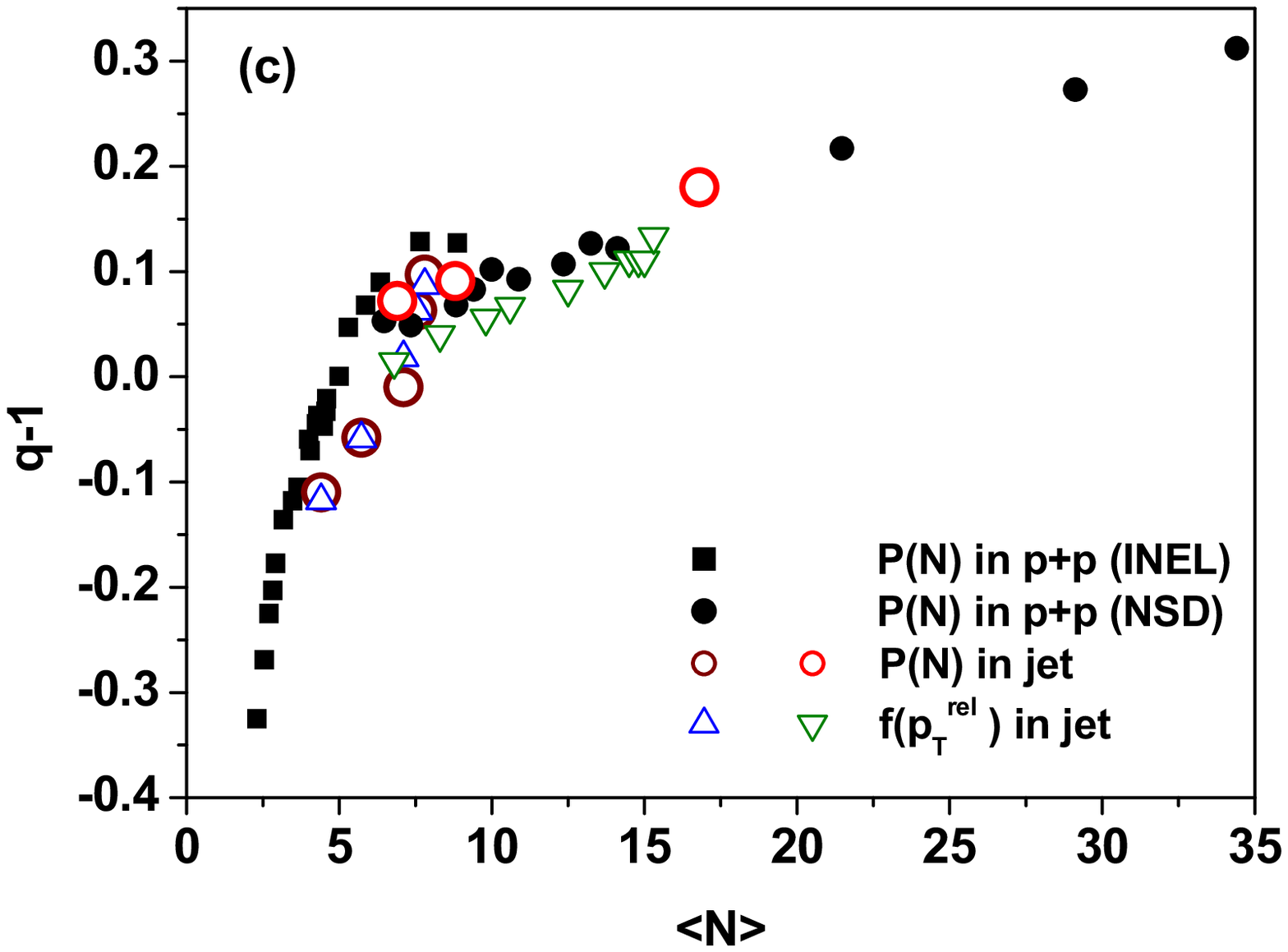}
\end{center}
\caption{(Color online) $(a)$ Transverse momentum spectra $p_T$  for jets (open symbols) compared with $p_T$ spectra for hadrons (full symbols) from proton-proton collisions at $\sqrt{s}= 7$ TeV (with arbitrary normalization at $p_T = 20$ GeV; data are from \cite{ATLAS-a,ATLAS-b,CMS-1,CMS-2}). The dashed line shows the fit using Eq. $(\ref{CT})$.
$(b)$ The power index $n$ extracted from fits to the jet spectra shown in $(a)$ (full symbols) compared with $n$ from $p_T$ distributions of charged particles (open symbols, data are from \cite{CMS-1,CMS-2,WWCT}.
$(c)$ Compilation of values of the parameters $q$ obtained from the $p_T$  spectra (triangles) and from the multiplicity distributions (circles). Triangles at small $\langle N \rangle$ are obtained from \cite{ATLAS-a,ATLAS-b}, those for larger $\langle N\rangle$ from \cite{ATLAS-1}. Full squares and circles are from data on multiparticle production in $p + p$ collisions: squares (inelastic data) are from the compilation for LAB energy $3.7 - 303$ GeV presented in \cite{AW}, full circles (non-single diffractive data) are from the compilation in \cite{G-G}, and open-red circles are from \cite{ATLAS-2}.
}
\label{F1}
\end{figure}

As will be seen, the first phenomenon is connected with the fact that large $p_T$ distributions follow a quasi power-like pattern which is best described by the Tsallis distribution \cite{Tsallis,Tsallis-b,WW_1}:
\begin{equation}
f\left( p_T\right) = C \left( 1 + \frac{p_T}{nT}\right)^{-n},\qquad n=\frac{1}{q-1}.\label{CT}
\end{equation}
This is a quasi power-law distribution with two parameters: power index $n$ (connected with the nonextensivity parameter $q$) and scale parameter $T$ (in many applications identified with temperature) \cite{WW_1}. In Fig. \ref{F1} we present examples of applications of the nonextensive approach to multiparticle distributions represented by Eq. (\ref{CT}). Panels $(a)$ and $(b)$ show the high quality Tsallis fit and also a kind of self-similarity of $p_T$ distributions of jets and hadrons \cite{WW_a}. Panel $(c)$ demonstrates that the values of the nonextensivity parameters $q$ for particles in jets correspond rather closely to values of $q$ obtained from the inclusive distributions measured in $pp$ collisions \cite{WW_a,WW_b}\footnote{This observation should be connected with the fact that multiplicity distributions, $P(N)$, are closely connected with the nonextensive approach, and that $q-1 = Var(N)/<N>^2 - 1/<N>$ for Negative Binomial, Poissonian and Binomial distributions \cite{WW_b}.}. In general, what one observes is the self-similar character of the production process in both cases originating from their cascading character, which always results in a Tsallis distribution\footnote{In fact, this is very old idea, introduced by Hagedorn in \cite{Hagedorn,Hagedorn-a}, that the production of hadrons proceeds through the formation of {\it fireballs which are a statistical equilibrium of an undetermined number of all kinds of fireballs, each of which in turn is considered to be a fireball}. This idea returned recently in the form of thermofractals introduced in \cite{AD}. Note that the Negative Binomial Distribution discussed below also has a self-similar  character \cite{SSofNBD}. Note also that the self-similarity  and fractality features of the multiparticle production process were exhaustively discussed in \cite{DeWolf,Kittel}.}. In the pure dynamical QCD approach to hadronization one could think of partons fragmenting into final state hadrons through the multiple sub-jet production \cite{Kittel}.

\section{Log-periodic oscillations in data on large $p_T$ momenta distributions}
\label{sec:LPO}

\begin{figure}[h]
\begin{center}
\includegraphics[scale=0.333]{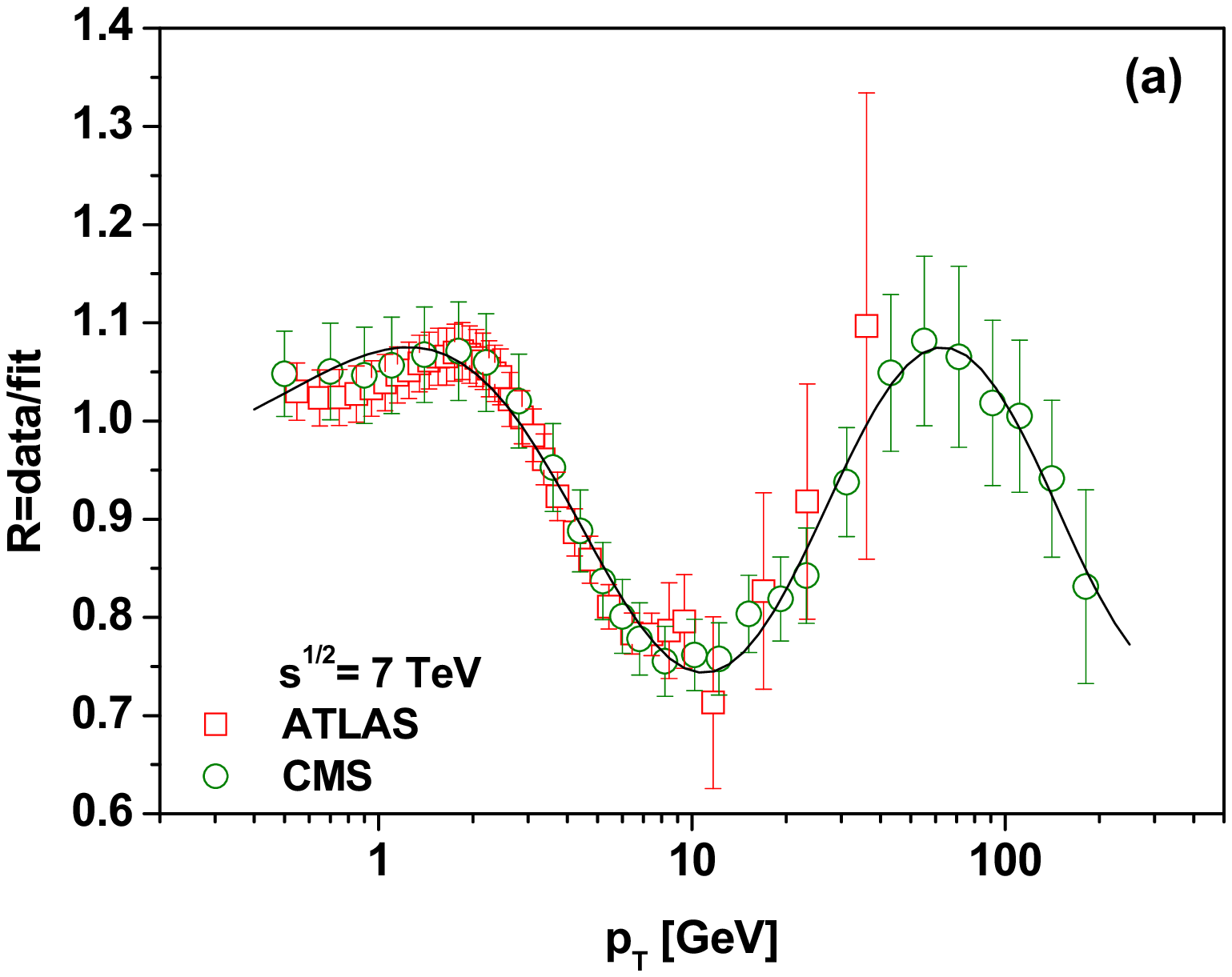}
\includegraphics[scale=0.333]{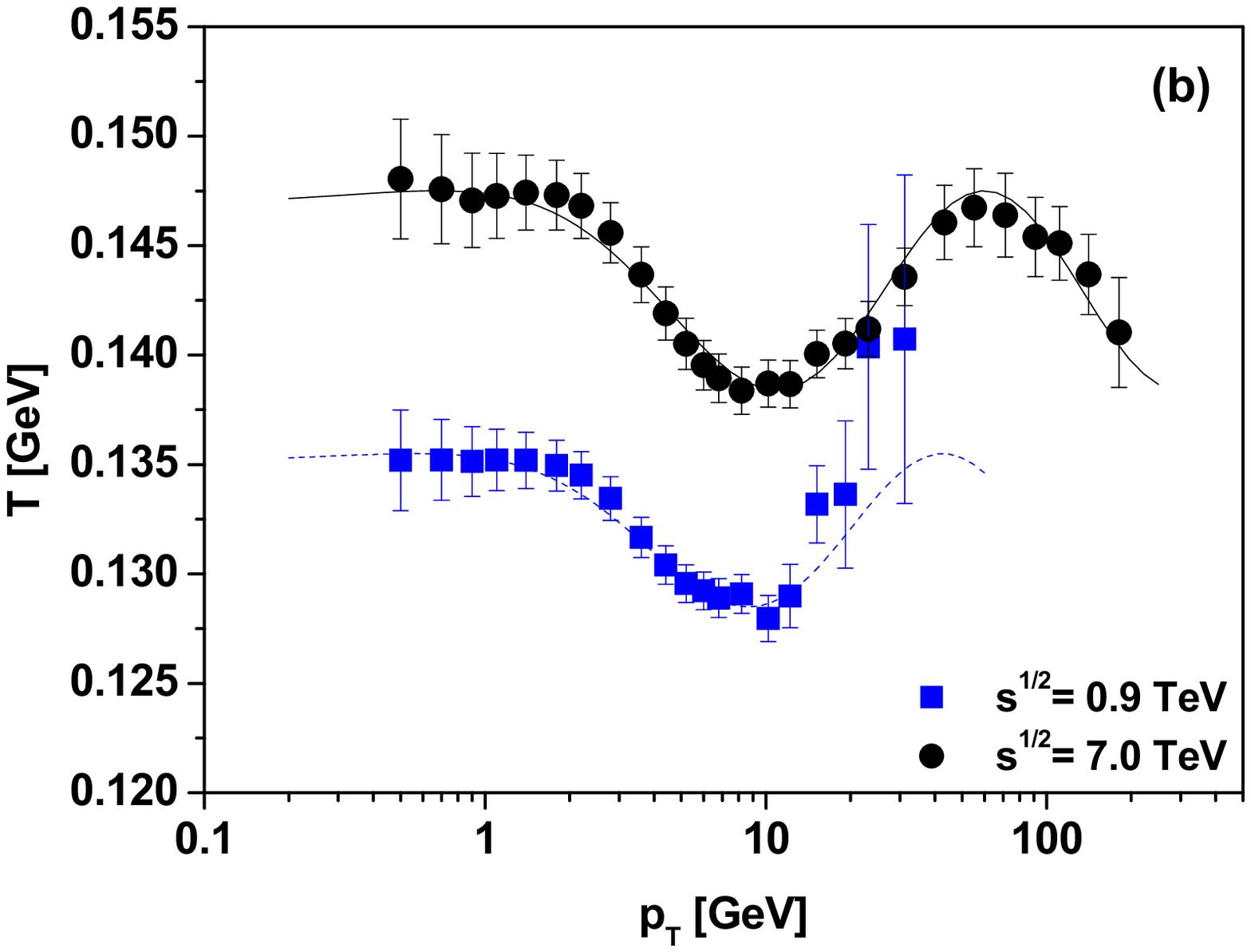}
\includegraphics[scale=0.333]{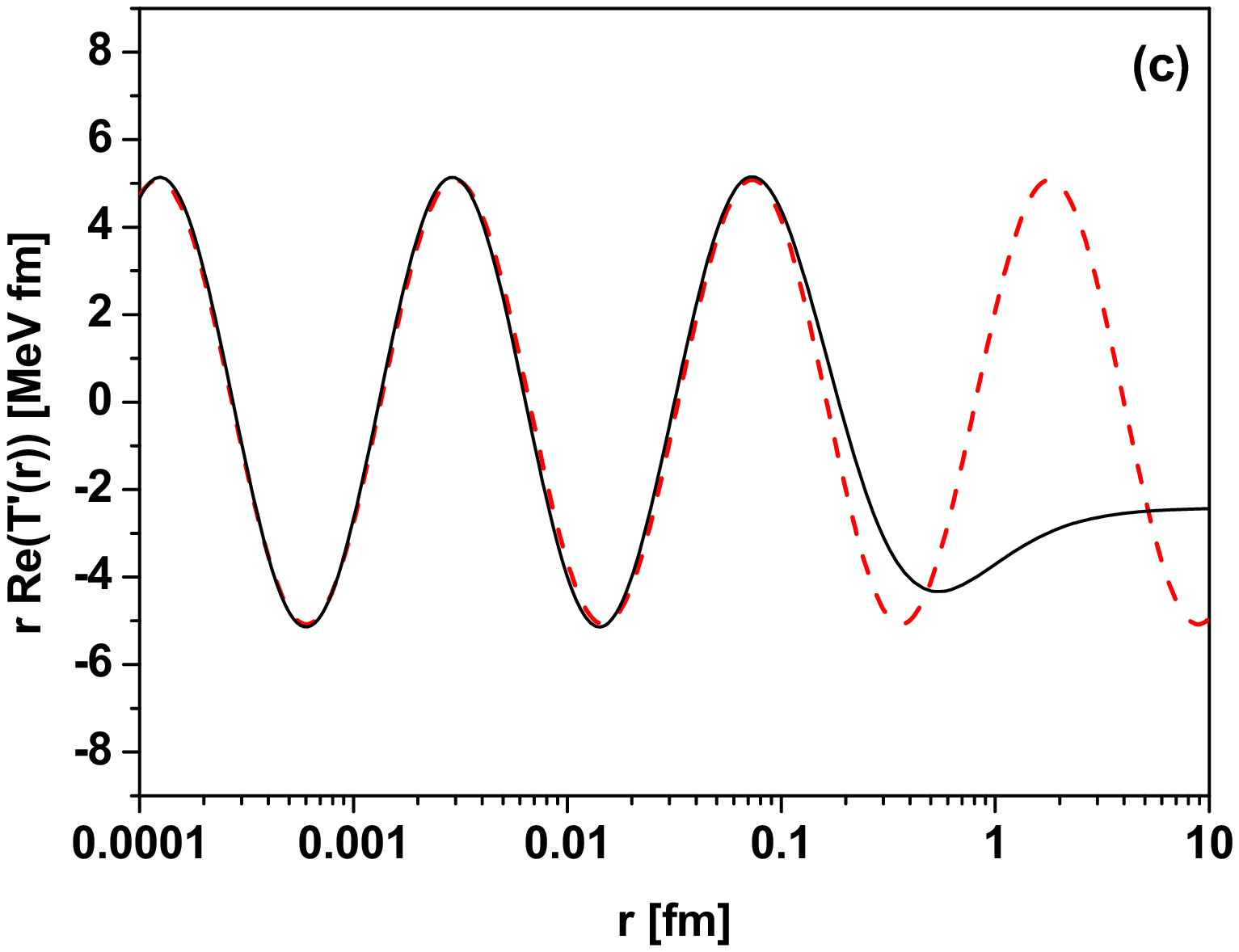}
\end{center}
\caption{(Color online)  $(a)$ The $R = data/fit$ ratio for $pp$ collisions at TeV from the CMS \cite{CMS-1,CMS-2,CMS-3} and ATLAS \cite{ATLAS-3} experiments fitted by Eq. (\ref{Rp_T}) with parameters: $a = 0.909$, $b = 0.166$, $c = 1.86$, $d = 0.948$ and $f = -1.462$. $(b)$ The log-periodic oscillations of $T = T\left( p_T\right)$ fitting $R\left( p_T\right)$ from panel $(a)$  for $pp$ collisions at $0.9$ and $7$ TeV from the CMS experiment \cite{CMS-1,CMS-2,CMS-3}.  They can be fitted by Eq. (\ref{Tp_T}) with parameters: $\tilde{a} = 0.132$, $\tilde{b} = 0.0035$, $\tilde{c} = 2.2$, $\tilde{d} = 2.0$, $\tilde{f} = -0.5$ for $0.9$ TeV and $\tilde{a} = 0.143$, $\tilde{b} = 0.0045$, $\tilde{c} = 2.0$, $\tilde{d} = 2.0$, $\tilde{f} = -0.4$ for $7$ TeV. $(c)$ The results of the Fourier transform of $T (p_T )$  from panel $(b)$ of Fig. \ref{F2}. The continuous line represents $rT'(r)$ versus $r$ and the dashed line denotes the function $rT'(r) = 5.1 \sin[(2\pi/3.2) \ln(1.24r)]$ fitting it for small values of $r$.
}
\label{F2}
\end{figure}

To start with the first example we note that, despite the exceptional quality of the Tsallis fit presented in panel $(a)$ of Fig. \ref{F1} (in fact, only such a two-parameter quasi-power like formula can fit the data over the whole range of $p_T$), the ratio $R = data/fit$ (which is expected to be flat a function of $p_T$, $R\left(p_T\right) \sim 1$) presented in panel $(a)$ of Fig. \ref{F2}, shows clear log-periodic oscillatory behaviour as a function of the transverse momentum $p_T$. In fact, it turns out that such behaviour occurs (at the same strength) in data from all LHC experiments, at all energies (provided that the range of the measured $p_T$ is large enough), and that its amplitude increases substantially for the nuclear collisions \cite{WW_c}. These observations strongly suggest that closer scrutiny should be undertaken to understand its possible physical origin.

Note that we have two parameters in the Tsallis formula, Eq. (\ref{CT}), $n$ and $T$, and each of them (or both, but we shall not consider such a situation) could be {\it a priori} responsible for the observed effect. We start with the power index $n$. In this case the observed oscillations may be related to some scale invariance present in the system and are an immanent feature of any power-like distribution \cite{Scaling}. In \cite{WW_2} we showed that they also appear in quasi power-like distributions of the Tsallis type. In general, they are attributed to a discrete scale invariance (connected with a possible fractal structure of the process under consideration) and are described by introducing a complex power index $n$ (or $q$). This, in turn, has a number of interesting consequences \cite{RWW,WW_3,WW_4}  like a complex heat capacity of the system or complex probability and complex multiplicative noise, all of them known already from other branches of physics. In short, one relies  on the fact that power like distributions, say $O(x) = C x^{-m}$, exhibit scale invariant behaviour, $Q(\lambda x) = \mu O(x)$, where parameters $\lambda$ and $\mu$ are, in general, related by the condition that $\mu \lambda^m = 1 = \exp(2\pi i k),~~k=0,~1,~2,\dots$,which means that the power index $m$ can take complex values: $m = - \frac{\ln \mu}{\ln \lambda} + i\frac{2\pi k}{\ln \lambda}$. In the case of a Tsallis distribution this means that Eq. (\ref{CT}) is decorated by some oscillating factor $R\left(p_T\right)$, the form of which (when keeping only the $k=0$ and $k=1$ terms) is:
\begin{equation}
R\left( p_T\right) = a + b \left[ \cos\left( p_T + d\right) + f\right] \label{Rp_T}
\end{equation}
As one can see in Fig. \ref{F2} it fits perfectly the observed log-oscillatory pattern.

The second possibility is to keep $n$ constant but allow the scale parameter $T$ to vary with $p_T$ in such a way as to allow a fit to the data on $R\left( p_T\right)$ \cite{WW_4}. The result is shown in panel $(b)$ of Fig. \ref{F2}. The resulting $T\left( p_T\right)$ has the form of log-periodic oscillations in $p_T$, which can be parameterized by
\begin{equation}
T\left( p_T\right) = \tilde{a} + \tilde{b} \left[ \sin\left( p_T + \tilde{d}\right) + \tilde{f}\right] \label{Tp_T}
\end{equation}
Such behaviour of $T\left( p_T\right)$ can originate from the well known stochastic equation for the temperature evolution which in the Langevin formulation has the form:
\begin{equation}
\frac{dT}{dt} + \frac{1}{\tau} T + \xi(t) = \Phi \label{LE}
\end{equation}
where $\tau$ is the relaxation time and $\xi(t)$ is time-dependent white noise. Assuming additionally that we have time-dependent transverse momentum, $p_T = p_T(t)$, increasing in ta way following the scenario of the preferential growth of networks \cite{WW_d},
\begin{equation}
\frac{dp_T}{dt} = \frac{1}{\tau_0}\left( \frac{p_T}{n} \pm T\right) \label{PrefG}
\end{equation}
($n$ is the power index and $\tau_0$ is some characteristic time step) one may write
\begin{equation}
\frac{1}{\tau_0}\left( \frac{p_T}{n} \pm T\right) \frac{d T}{ d p_T} + \frac{1}{\tau} T + \xi(t) = \Phi. \label{LEpT}
\end{equation}
Eq. (\ref{Tp_T}) is obtained in two cases: $(i)$ the noise term increases logarithmically with $p_T$ while the relaxation time $\tau$ remains constant:
\begin{equation}
\xi\left(t, p_T\right) = \xi_0(t) + \frac{\omega^2}{n}\ln \left( p_T\right), \label{p_T_increase}
\end{equation}
$(ii)$ the white noise is constant, $\xi\left( t, p_T\right) = \xi_0(t)$, but the relaxation time becomes $p_T$-dependent, for example
\begin{equation}
\tau = \tau\left( p_T\right) = \frac{n\tau_0}{n + \omega^2 \ln\left( p_T\right)} \label{Tau_dep}
\end{equation}
(in both cases $\omega$ is some new parameter \cite{WW_1,WW_4})\footnote{To fit data one needs only a rather small admixture of the stochastic processes with noise depending on $p_T$. The main contribution comes from the usual energy-independent Gaussian white noise. Note that whereas each of the proposed approaches is based on a different dynamical picture, they are numerically equivalent.}.

However, we can use Eq. (\ref{Tp_T}) in a way that allows us to look deeper into our dynamical process. To this end we calculate the Fourier transform of $T\left( p_T\right)$ presented there,
\begin{equation}
T(r) = \sqrt{\frac{2}{\pi}}\int^{\infty}_0 \, T\left( p_T\right) e^{ip_T r} dp_T = T_0 + T'(r).   \label{FT}
\end{equation}
One can see now how the temperature $T$ (more exactly, its part which varies with $r$, $T'(r)$), changes with distance $r$ from the collision axis (defined in the plane perpendicular to the collision axis and located at the collision point). The result of this operation is seen in panel $(c)$ of Fig. \ref{F2} as specific log-periodic oscillations in $p_T$. Such behaviour can be studied by considering the flow of a compressible fluid in a cylindrical source. Assuming oscillations with small amplitude and velocity $\boldsymbol{v}$, and introducing the velocity potential $f$ such that $\boldsymbol{v} = \boldsymbol{grad} f$, one finds that $f$ must satisfy the following cylindrical wave equation:
\begin{equation}
\frac{1}{r} \frac{\partial}{\partial r}\left( r \frac{\partial f}{\partial r}\right) - \frac{1}{c^2}\frac{\partial^2 f}{\partial t^2} =0.    \label{partialwe}
\end{equation}
It can be shown that this represents a travelling sound wave with velocity $\boldsymbol{v}$ in the direction of propagation. Because the oscillating part of the temperature, $T'$, is related to the velocity $v$,
\begin{equation}
T' = \frac{c \kappa T}{c_P} v, \label{ResLandau}
\end{equation}
where $\kappa = \frac{1}{V}\left( \frac{\partial V}{\partial T}\right)_P$ is the coefficient of thermal expansion and $c_P$ denotes the specific heat  at constant pressure  \cite{Landau}, in the case of a monochromatic wave where $f(r,t) = f(r)\exp(-i\omega t)$, we have that
\begin{equation}
\frac{\partial^2 f(r)}{\partial r^2} + \frac{1}{r} \frac{\partial f(r)}{\partial r} + K^2 f(r) = 0,\qquad \qquad
K=K(r) = \frac{\omega}{c(r)},  \label{eqn}
\end{equation}
with $K$ being the wave number depending in general on $r$. For
\begin{equation}
K(r) = \frac{\alpha}{r} \label{Kr}
\end{equation}
the solution of Eq. (\ref{eqn}) takes the form of some log-periodic oscillation,
\begin{equation}
f(r) \propto \sin [ \alpha \ln(r)]. \label{solution}
\end{equation}
Because in our case $f(r) \propto v r$, using Eq. (\ref{ResLandau}), we can write that
\begin{equation}
rT'(r) \propto \frac{c\kappa T_0}{c_P}f(r) = \frac{c \kappa T_0}{c_P} \sin[ \alpha \ln (r)], \label{final}
\end{equation}
which is what we have used in describing the $T'(r)$ presented in panel $(c)$ of Fig. \ref{F2}. The space picture of the collision (in the plane perpendicular to the collision axis and located at the collision point) which emerges is some regular logarithmic structure for small distances which disappears when $r$ reaches the dimension of the nucleon, i.e., for $r \sim 1$ fm. Whether it is connected with the parton structure of the nucleon remains for the moment an open question.

To end this section note that Eq. (\ref{eqn}) with $K(r)$ given by Eq. (\ref{Kr}) is {\it scale invariant} and that $f(\lambda r) = f(r)$. Note also that in the variable $\xi = \ln r$ Eq. (\ref{eqn}) is also {\it self-similar} because in this variable it takes the form of the {\it traveling wave equation}
\begin{equation}
\frac{\partial ^2 F(\xi)}{\partial \xi^2}  +  \alpha^2 F(\xi ) = 0 \label{Fxi}
\end{equation}
which has the self-similar solution
\begin{equation}
F(\xi) \propto \cos[\alpha \xi],  \label{solFxi}
\end{equation}
for which $F(\xi + \ln \lambda) = F(\xi)$ (with $\alpha = \frac{2\pi k}{\ln \lambda}$ or $\lambda = \exp\left( \frac{2\pi k}{\alpha}\right)$ where $k = 1, 2, 3,\dots$). This is the so called {\it self-similar solution of the second kind} usually encountered in the description of the intermediate asymptotic. Such asymptotics are observed in phenomena which do not depend on the initial conditions because sufficient time has already passed, nevertheless the system considered is still out of equilibrium \cite{BB}.

\section{Oscillations hidden in the multiplicity distributions data}
\label{sec:PN}

Whereas the previous section was concerned with oscillations hidden in the distributions of produced particles in transverse momenta, $f\left(p_T\right)$ (which by using the Fourier transformation allowed us to gain some insight into the space picture of the interaction process), in this section we shall concentrate on another important characteristic of the  multiparticle  production process, namely on the question of how many particles are produced and with what probability, i.e., on the multiplicity distribution function, $P(N)$, where $N$ is the observed number of particles. It is usually one of the first observables measured in any multiparticle production experiment \cite{Kittel}.

At first we note that any $P(N)$ can be defined in terms of some recurrence relation, the most popular takes the form:
\begin{equation}
(N+1)P(N+1) = g(N)P(N)\qquad {\rm where}\qquad g(N) = \alpha + \beta N. \label{RR}
\end{equation}
Such a linear form of $g(N)$ leads to a Negative Binomial Distribution (NBD), Binomial Distribution (BD) or Poisson Distribution (PD):
\begin{eqnarray}
\hspace{-15mm} NBD: && P(N) = \frac{\Gamma(N+k)}{\Gamma(N+1)\Gamma(k)} p^N (1 - p)^k
                               \quad \hspace{16mm} {\rm with}\quad \alpha = kp,\hspace{8mm} \beta = \frac{\alpha}{k} \label{NBD}\\
\hspace{-11mm} BD: && P(N) = \frac{K!}{N!(K - N)!} p^N (1 - p)^{K-N}\quad
     \hspace{12mm}{\rm with}\quad \alpha = \frac{Kp}{1-p}, ~~~\beta = -\frac{\alpha}{K},
\label{BD}\\
\hspace{-15mm}PD: && P(N) = \frac{\lambda^N}{N!} \exp( - \lambda)\quad \hspace{38mm} {\rm with}\quad  \alpha = \lambda, \hspace{8mm} ~~~\beta = 0. \label{PD}
\end{eqnarray}
Suitable modifications of $g(N)$ result in more involved distributions $P(N)$ (cf. \cite{JPG} for references).

The most popular form of $P(N)$ is the NBD type of distribution, Eq. ({\ref{NBD}). However, with growing energy and  number of produced secondaries the NBD starts to deviate from data for large $N$ and one has to use combinations of NBDs (two \cite{GU}, three \cite{Z} and multi-component NBDs \cite{DN} were proposed) or try to use some other form of $P(N)$ \cite{Kittel,DG,G-OR}. For example, in Fig. \ref{F3} $(a)$ a single NBD is compared with $2$-NBD. However, as shown there, the improvement, although substantial, is not completely adequate. It is best seen when looking at the ratio $R = P_{CMS}(N)/P_{fit}(N)$ which still shows some wiggly structure (albeit substantially weaker than in the case of using only a single NBD to fit the data). Taken seriously, this observation suggests that there is some additional information hidden in the $P(N)$. The question of how to retrieve this information was addressed in \cite{JPG} by resorting to a more general form of Eq. (\ref{RR}), usually used in counting statistics when dealing with cascade stochastic processes \cite{CSP} in which all multiplicities are connected.
\begin{figure}[h]
\begin{center}
\includegraphics[scale=0.333]{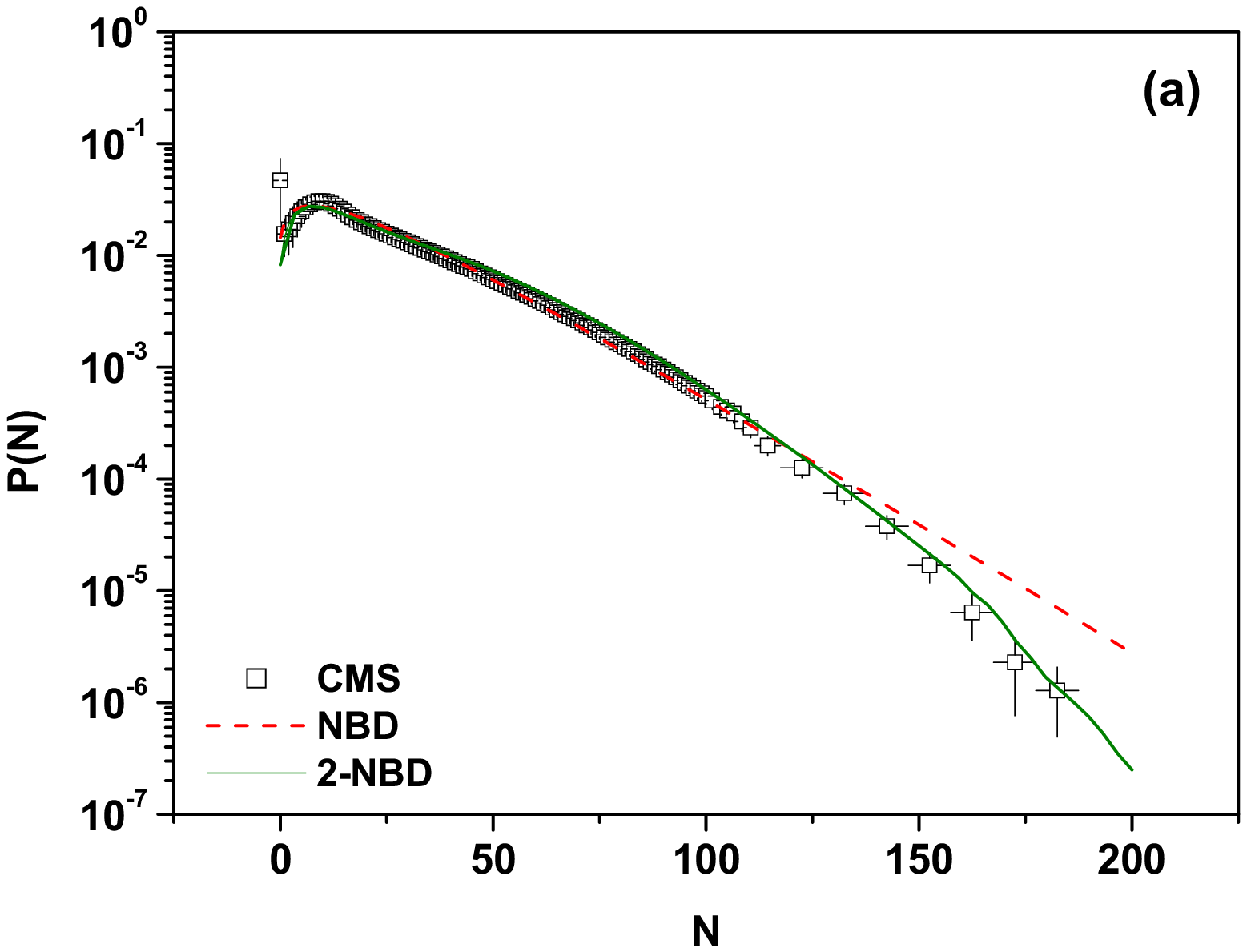}
\includegraphics[scale=0.333]{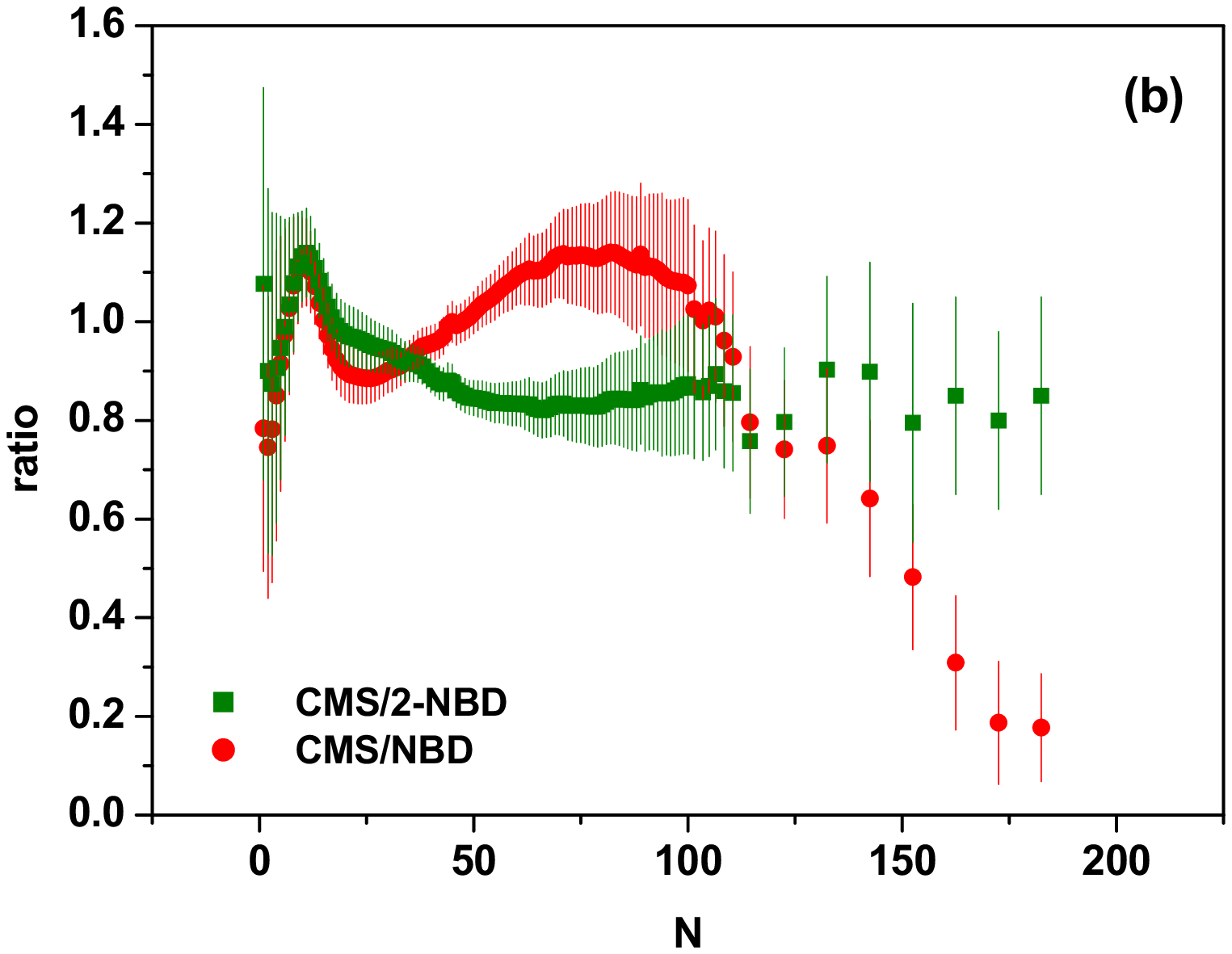}
\includegraphics[scale=0.333]{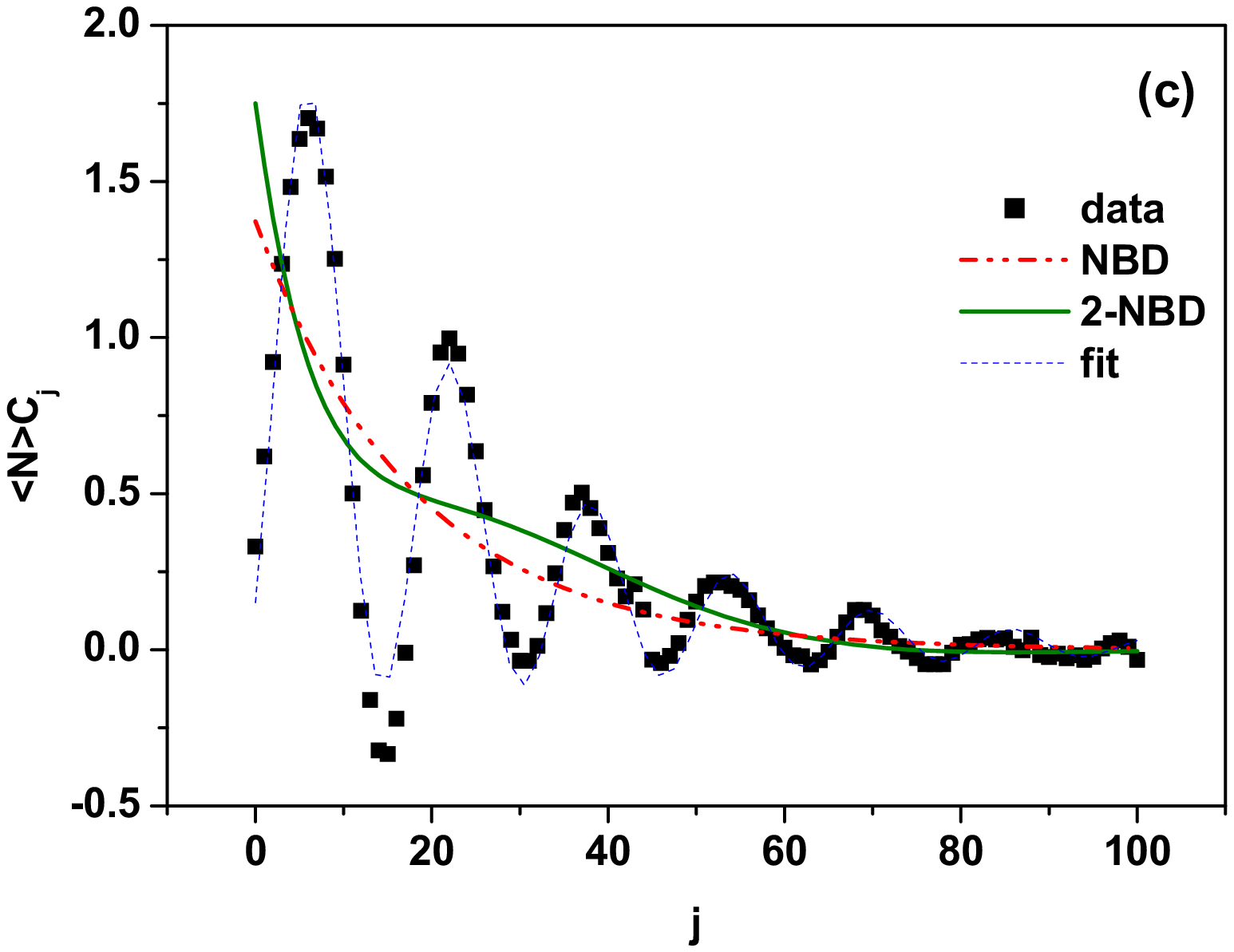}
\end{center}
\caption{(Color online) $(a)$ Charged hadron multiplicity distributions for the pseudorapidity range $|\eta| < 2$ at $\sqrt{s} = 7$ TeV, as given by the CMS experiment \cite{CMS-4} (points), compared with the NBD for parameters $\langle N\rangle = 25.5$ and $k = 1.45$ (dashed line) and with the $2$-component NBD (solid line) with parameters from \cite{NBD-PG}. $(b)$ Multiplicity dependence of the ratio $R = P_{CMS}(N)/P_{fit}(N)$ for the NBD (red circles) and for the $2$-component NBD for the same data as in panel $(a)$ (green squares). $(c)$ Coefficients $C_j$ emerging from the data and NBD fits presented in panel $(a)$. The data points are fitted by Eq. (\ref{CjFit}), see text for details.
}
\label{F3}
\end{figure}
In this case one has coefficients $C_j$ defining the corresponding $P(N)$ in the following way:
\begin{equation}
(N + 1)P(N + 1) = \langle N\rangle \sum^{N}_{j=0} C_j P(N - j). \label{Cj}
\end{equation}
These coefficients contain the memory of particle $N+1$ about all $N-j$ previously produced particles. Assuming now that all $P(N)$ are given by experiment one can reverse Eq. (\ref{Cj}) and obtain a recurrence formula for the coefficients $C_j$,
\begin{equation}
\langle N\rangle C_j = (j+1)\left[ \frac{P(j+1)}{P(0)} \right] - \langle N\rangle \sum^{j-1}_{i=0}\left[ \frac{P(j-i)}{P(0)} \right]. \label{rCj}
\end{equation}
As can be seen in Fig. \ref{F3} $(c)$ the coefficients $C_j$ obtained from the data presented in Fig. \ref{F3} $(a)$ show oscillatory behaviour (with period roughly equal to $16$) gradually disappearing with $N$. They can be fitted by the following formula:
\begin{equation}
\langle N\rangle C_j = \left( a^2 + b^2\right)^{j/2} \sin\left[ c + j \arctan(b/a)\right] +d^j, \label{CjFit}
\end{equation}
with parameters: $a = 0.89$, $b = 0.37$, $c = 5.36$, $d = 0.95$. Such oscillations do not appear in the single NBD fit presented in Fig. \ref{F3} $(a)$ and there is only a small trace of oscillations for the $2$-NBD fit presented in Fig. \ref{F3} $(a)$. This is because for a single NBD one has a smooth exponential dependence of the corresponding $C_j$ on the rank $j$,
\begin{equation}
C_j = \frac{k}{\langle N\rangle} p^{j+1} = \frac{k}{k + m} \exp( j \ln p), \label{CjNBD}
\end{equation}
and one can expect any structure only for the multi-NBD cases \cite{JPG}.

Before proceeding further let us note that the coefficients $C_j$ are closely related to the so called {\it combinants} $C^{\star}_j$ which were introduced in \cite{KG} (see also \cite{Kittel,CombUse,H}) and are defined in terms of the generating function $G(z)$ as
\begin{equation}
C^{\star}_j = \frac{1}{j!} \frac{d^j \ln G(z)}{d z^j}\bigg|_{z=0},\qquad {\rm where}\qquad G(z) = \sum^{\infty}_{N=0} P(N) z^N , \label{CombDef-1}
\end{equation}
or by the relation \cite{JPG},
\begin{equation}
\ln G(z) = \ln P(0) + \sum^{\infty}_{j=1} C^{\star}_j z^j. \label{CombDef-2}
\end{equation}
From the above one can deduce that \cite{JPG}
\begin{equation}
C_j = \frac{j+1}{\langle N\rangle} C^{\star}_{j+1}. \label{connection}
\end{equation}
This means that one can rewrite the recurrence  relation, Eq. (\ref{Cj}),  in terms of the combinants $C_j^{\star}$:
\begin{equation}
(N + 1)P(N + 1) = \sum^{N}_{j=0} (j+1)C_j^{\star} P(N - j). \label{Cj_star}
\end{equation}
When compared with Eq.(\ref{Cj}) this allows us to express our coefficients $C_j$, which henceforth we shall call {\it modified combinants}, by the generating function $G(z)$ of $P(N)$:
\begin{equation}
\langle N\rangle C_j = \frac{1}{j!} \frac{ d^{j+1} \ln G(z)}{d z^{j+1}}\bigg|_{z=0}. \label{GF_Cj}
\end{equation}
This is the relation we shall use in what follows when calculating the $C_j$ from distributions defined by some $G(z)$.

To continue our reasoning, note first that whereas a single NBD does not lead to oscillatory behaviour of the modified combinants, $C_j$, there is a distribution for which the corresponding $C_j$ oscillate in a maximum way. This is the BD for which the modified combinants are given by the formula
\begin{equation}
C_j = (-1)^j \frac{K}{\langle N\rangle} \left( \frac{\langle N\rangle}{K-\langle N\rangle}\right)^{(j+1)} = \frac{(-1)^j}{1 - p}\left( \frac{p}{1 - p}\right)^{j}, \label{C_jBD}
\end{equation}
which oscillates rapidly with a period equal to $2$. In Fig. \ref{F4} $(a)$ one can see that the amplitude of these oscillations depends on $p$, generally the $C_j$ increase with rank $j$ for $p > 0.5$ and decrease for $p < 0.5$. However, their general shape lacks the fading down feature of the $C_j$ observed experimentally. This suggests that the BD alone is not enough to explain the data but must be somehow combined with some other distribution\footnote{In fact, in \cite{BSWW} we have already used a combination of {\it elementary emitting cells (EEC)} producing particles following a geometrical distribution (our aim at that time was to explain the phenomenon of Bose-Einstein correlations). For constant number $k$ of {\it EECs} one gets the  NBD as the resultant $P(N)$, whereas for $k$ distributed according to the BD, the resulting $P(N)$ was a modified NBD. However, we could not find at present a set of parameters providing both the observed $P(N)$ and oscillating $C_j$. Note that originally the NBD was seen as a compound Poisson distribution, with the number of clusters given by a Poissonian distribution and the particle inside the clusters distributed according to a logarithmic distribution \cite{GVH}}.}.

\begin{figure}[h]
\begin{center}
\includegraphics[scale=0.5]{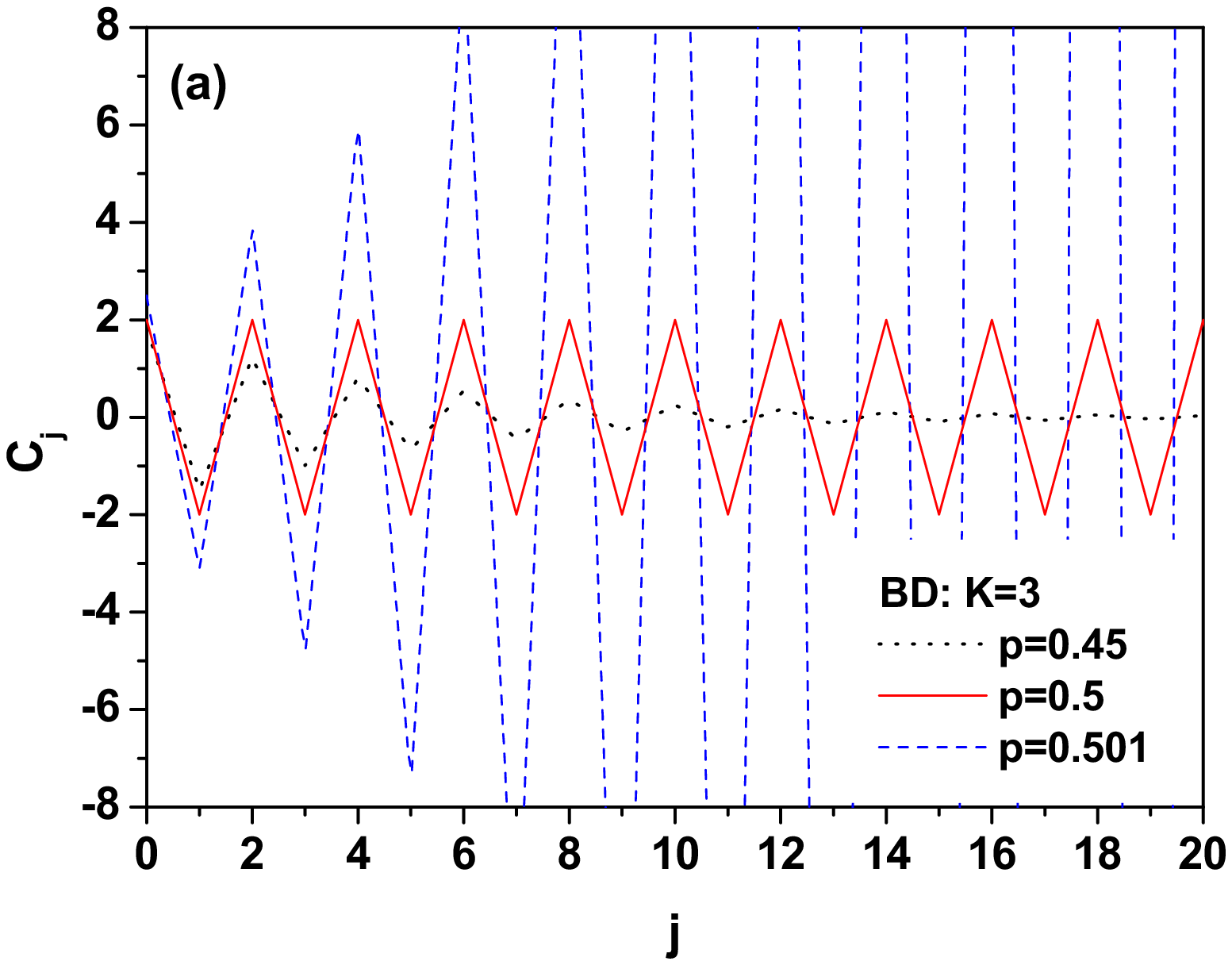}
\includegraphics[scale=0.5]{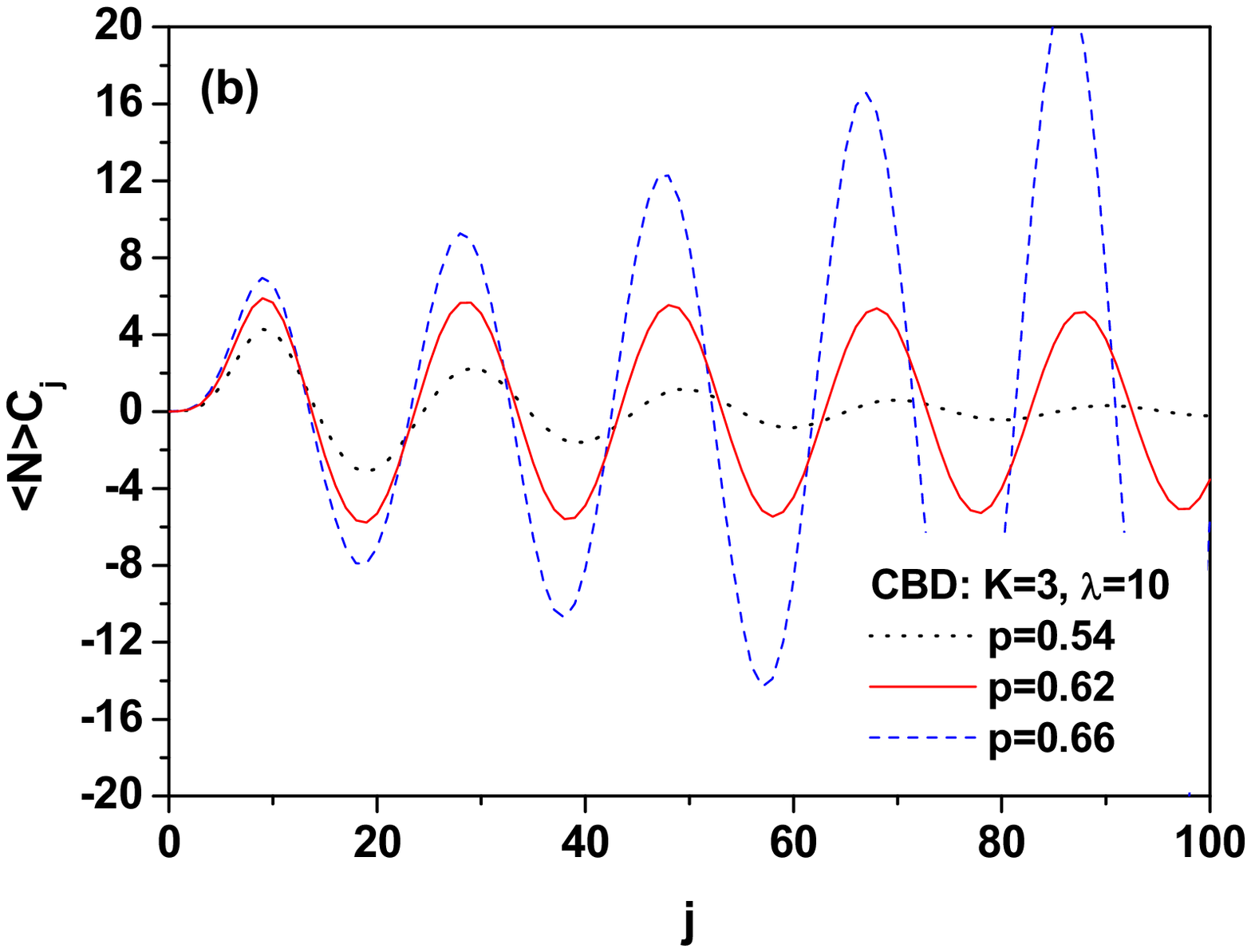}
\end{center}
\caption{(Color online) Examples of $C_j$ for Binomial Distributions $(a)$ (from Eq. (\ref{C_jBD})) and Compound Binomial Distributions $(b$) from Eq. (\ref{C_BD_PD}).
}
\label{F4}
\end{figure}

We resort therefore to the idea of {\it compound distributions} (CD) \cite{Compound} which, from the point of view of the physics involved in our case, could for example describe a production process in which a number $M$ of some objects (clusters/fireballs/etc.) is produced following, in general, some distribution $f(M)$ (with generating function $F(z)$), and subsequently decaying  independently into a number of secondaries, $n_{i = 1,\dots, M}$, always following some other (the same for all) distribution, $g(n)$ (with a generating function $G(z)$). The distribution $h(N)$, where
\begin{equation}
N = \sum_{i=0}^M n_i, \label{N}
\end{equation}
is a compound distribution of $f$ and $g$: $h = f\otimes  g$. For compound distributions we have that
\begin{equation}
\langle N\rangle = \langle M\rangle \langle n\rangle\qquad {\rm and}\qquad Var(N) = \langle M\rangle Var(n) + Var(M) \langle n\rangle^2 \label{CD_moments}
\end{equation}
and its generating function $H(z)$ is equal to,
\begin{equation}
H(z) = F[G(z)]. \label{CD_GF}
\end{equation}

It should be mentioned that for the class of distributions of $M$ that satisfy our recursion relation Eq. (\ref{RR}) the compound distribution $h=f\otimes  g$ is given by the so called Panjer's recursion relation \cite{Panjer},
\begin{equation}
Nh(N) = \sum^{N}_{j=1}[ \beta N + (\alpha - \beta)j ] g(j) h(N-j) = \sum^{N}_{j=1} C^{(P)}_j (N) h(N-j),
 \label{PP}
\end{equation}
with the initial value $h(0)=f(0)$. However, the coefficients $C^{(P)}_j$ occurring here depend on $N$, contrary to our recursion given by Eq. (\ref{Cj}) where the modified combinants, $C_j$, are independent of $N$. Moreover,  Eq. (\ref{Cj}) is not limited to the class of distributions satisfying Eq.(\ref{RR}) but is valid for any distribution $P(N)$. For this reason the recursion relation Eq. (\ref{PP}) is not suitable for us.

To visualize the compound distribution in action we take for $f$ a Binomial Distribution with generating function $F(z) = (pz + 1 - p)^K$, and for $g$ we take a Poisson distribution with generating function $G(z) = \exp[ \lambda (z-1)]$. The generating function of the resultant distribution is now equal to:
\begin{equation}
H(z) = \left\{ p \exp[ \lambda (z-1)] + 1 -p \right\}^K \label{H_FG}
\end{equation}
and the corresponding modified combinants are:
\begin{eqnarray}
\langle N\rangle C_j &=& \frac{K \lambda^{j+2} \exp(-\lambda)}{j!} \sum^{j+2}_{i=1} \left[ \frac{p}{1 - p + p \exp(-\lambda)}\right]^i \frac{1}{i}\sum^{i}_{k=0} (-1)^{k+1}\binom{i}{k} k^{j+1} = \nonumber\\
&=& \frac{K \lambda^{j+2} \exp(-\lambda)}{j!} \sum^{j+2}_{i=1}\left[ \frac{p}{1 - p + p \exp(-\lambda)}\right]^i S(j+1,i) \label{C_BD_PD}
\end{eqnarray}
where
\begin{equation}
S(n,k) = \left\{ \begin{array}{c}
n\\
k
\end{array}\right\} = \frac{1}{k!} \sum_{i=0}^{k} (-1)^{k-i} \binom{k}{i}i^n \label{S}
\end{equation}
\begin{figure}[t]
\begin{center}
\includegraphics[scale=0.51]{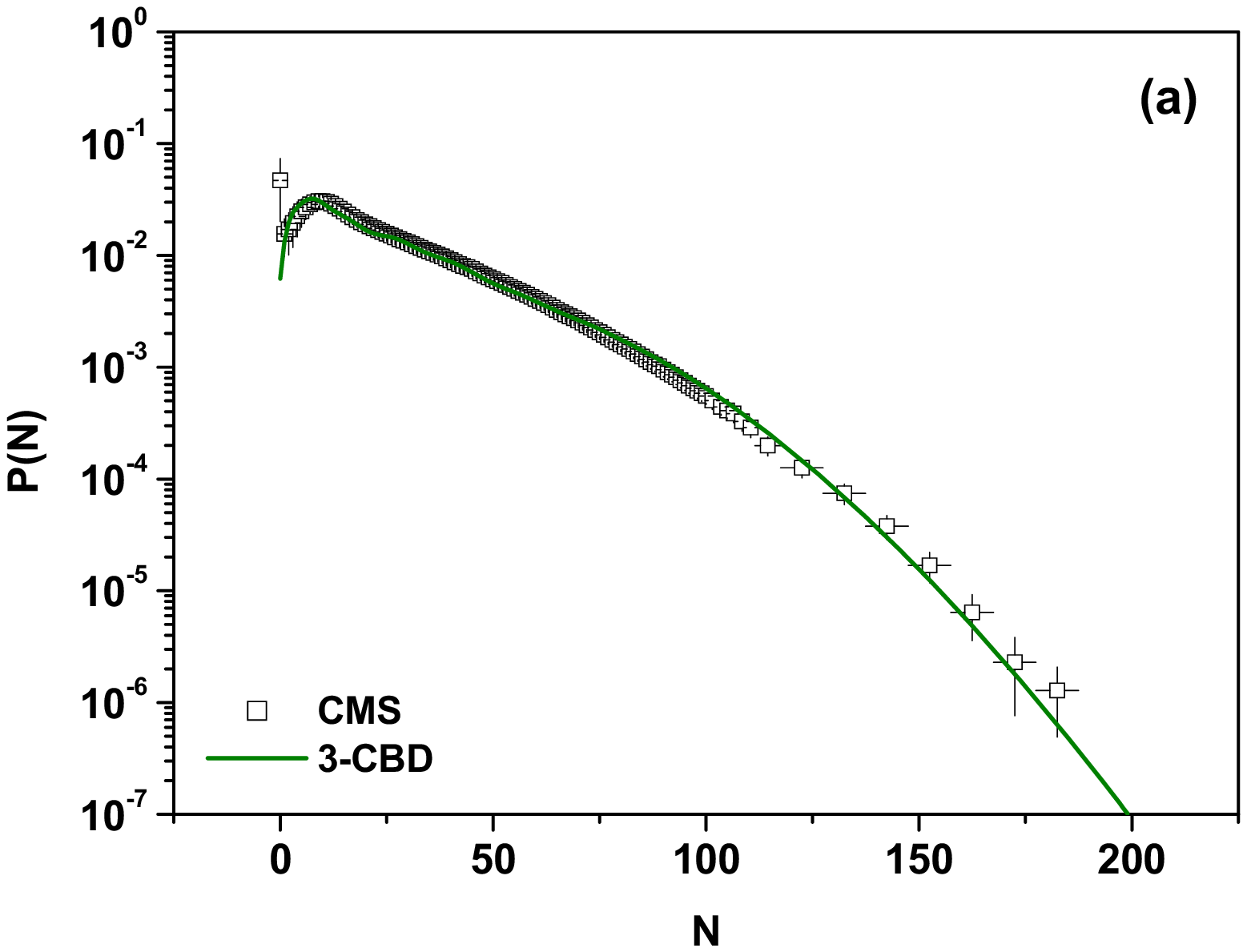}
\includegraphics[scale=0.51]{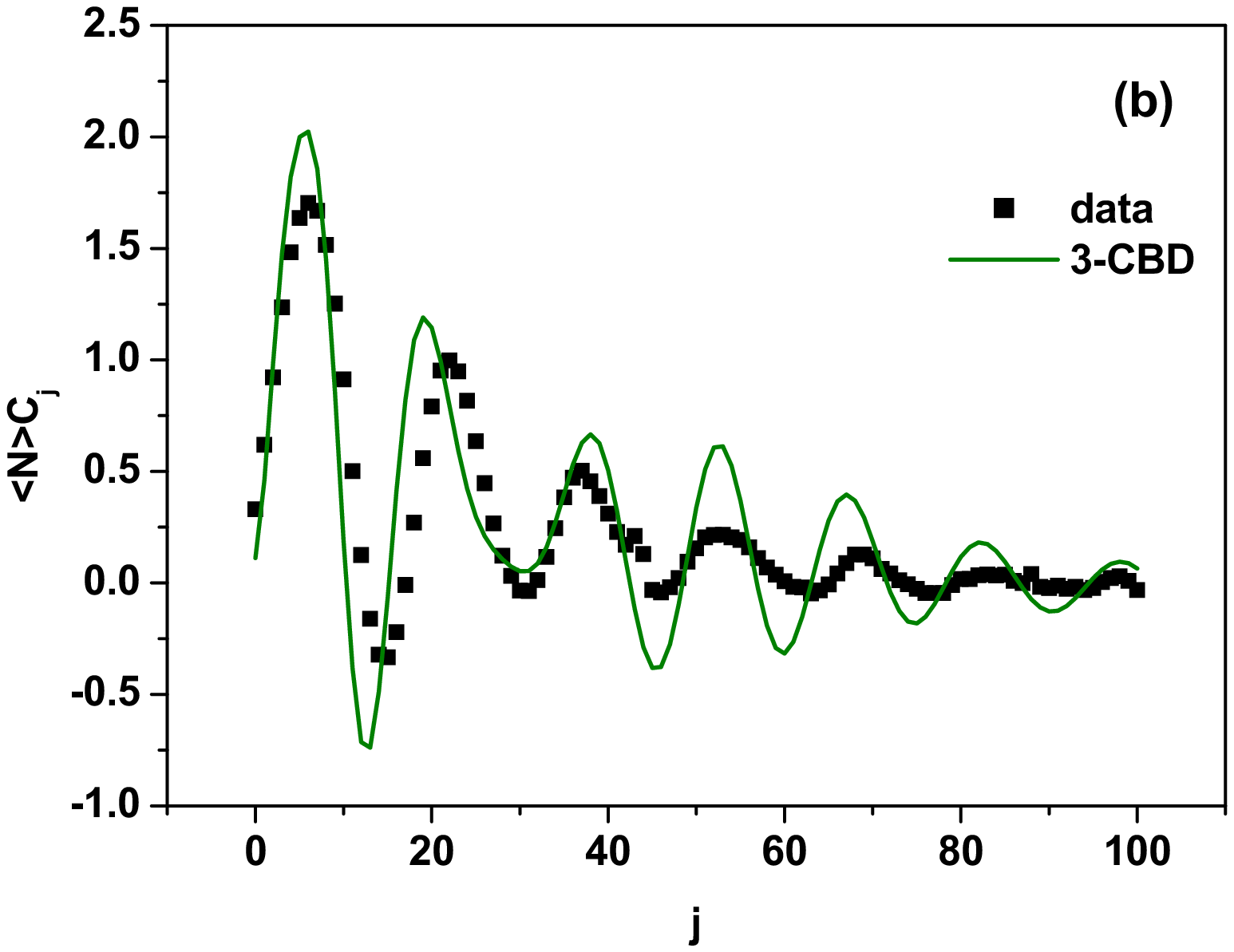}
\end{center}
\caption{(Color online) Results of using the compound distribution (CBD) approach given by Eq. (\ref{3CBD}) with parameters: $\omega_1 = 0.34$, $\omega_2 = 0.4$, $\omega_3 = 0.26$; $p_1 = 0.22$, $p_2 = 0.22$, $p_3 = 0.12$; $K_1 = 10$, $K_2 = 12$, $K_2 = 30$ and $\lambda_1 = 4$, $\lambda_2 = 9$, $\lambda_3 = 14$. $(a)$  Charged hadron multiplicity distributions for $|\eta| < 2$ at $\sqrt(s) = 7$ TeV, as given by the CMS experiment \cite{CMS-4} (points), compared with a $3$-component CBD, Eq. (\ref{3CBD}). $(b)$ Coefficients $C_j$ emerging from the CMS data used in panel $(a)$ compared with the corresponding $C_j$ obtained from the  $3$-component compound binomial distribution ($3$-CBD).
}
\label{F5}
\end{figure}
is the Stirling number of the second kind. Fig. \ref{F4} $(b)$ shows the above modified combinants for the Compound Binomial Distribution (CBD) (a combination of a BD with a PD) with $K=3$ and $\lambda =10$ calculated for three different values of $p$ in the BD: $p=0.54,~0.62,~0.66$. Note that in general the period of the oscillations is equal to $2\lambda$, i.e., in Fig. \ref{F4} $(b)$ where $\lambda = 10$ it is equal $20$. The multiplicity distribution in this case is
\begin{eqnarray}
P(0) &=& \left( 1 - p + p e^{-\lambda}\right)^K,\\   \label{P(0)}
P(N) &=& \frac{1}{N!} \frac{d^N H(z)}{dz^N}\bigg|_{z=0} = \frac{1}{N!}\sum^K_{i=1}i!\binom{K}{i}\left( \lambda p e^{-\lambda}\right)^i \left( 1 - p + p e^{-\lambda}\right)^{K-i} S(N,i) \label{P(N)}
\end{eqnarray}
(the proper normalization comes from the fact that $H(1)=1$). This shows that the choice of a BD as the basis of the CDs to be used  seems to be crucial to obtain oscillatory $C_j$ (for example, a compound distribution formed from a NBD and some other NBD provides smooth $C_j$).

Unfortunately, such a single component CBD (depending on three parameters: $p$, $K$ and $\lambda$, $P(N) = h(N;p,K,\lambda)$ does not describe the experimental $P(N)$. We return therefore to the idea of using a multicomponent version of the CBD, for example a $3$-component CBD defined as (with $w_i$ being weights):
\begin{equation}
P(N) = \sum_{i=1,2,3} w_i h\left(N; p_i, K_i, \lambda_i\right);\qquad \qquad \sum_{1=1,2,3} w_i = 1.  \label{3CBD}
\end{equation}
As can be seen in Fig. \ref{F5}, this time the fit to $P(N)$ is quite good and the modified combinants $C_j$ follow an oscillatory pattern as far as the period of the oscillations is concerned, albeit their amplitudes still decay too slowly.

\section{Summary}
\label{sec:SUM}

To summarize, we would like to mention that power law and quasi power-law distributions are ubiquitous in many different,  apparently very disparate, branches of science (like, for example, earthquakes, escape probabilities in chaotic maps close to crisis, biased diffusion of tracers in random systems, kinetic and dynamic processes in random quenched and fractal media, diffusion limited aggregates, growth models, or stock markets near financial crashes, to name only a few). Also ubiquitous is the fact that, in most cases, they are decorated with log-periodic oscillations of different kinds \cite{Scaling,WW_2,WW_3,WW_4}. It is then natural to expect that oscillations of certain variables constitute a universal phenomenon which should occur in a large class of stochastic processes, independently of the microscopic details. In this paper we have concentrated on some specific oscillation phenomena seen at LHC energies in transverse momentum distributions.  Their log-periodic character suggests that either the exponent of the power-like behavior of these distributions is complex, or that there is a scale parameter which exhibits specific log-periodic oscillations. Whereas the most natural thing seems to attribute the observed oscillations to some discrete scale invariance present in the system considered \cite{Scaling,WW_2}, it turns out that such scale invariant functions also satisfy some specific wave equations showing a self-similarity property \cite{WW_c}. In both cases these functions exhibit log-periodic behavior.

Concerning the second topic considered here, the presence of oscillations in counting statistics, one should realize that it is also a well established phenomenon. The known examples include oscillations of the high-order cumulants of the probability distributions describing the transport through  a double quantum dot, oscillations in quantum optics (in the photon distribution function in slightly squeezed states) (see \cite{PNAS} for more information and references). In elementary particle physics oscillations of the so called $H_q$ moments, which represent ratios of the cumulants to factorial moments, also have a long history \cite{Kittel,DG,G-OR}.

Our expectation that the oscillations discussed here could also be observed (and successfully measured) in multiparticle production processes is new. To see them one must first deduce from the experimental data on the multiplicity distribution $P(N)$ the so called {\it modified combinants} $C_j$ (which are defined by the recurrence relation presented in Eq. ({\ref{Cj}); note that, contrary to the $H_q$ moments, the $C_j$ are independent of the multiplicity distribution $P(N)$ for $N > j$). In the case when these modified combinants show oscillatory behavior,  they can be used to search for some underlying dynamical mechanism which could be responsible for it. The present situation is such that the measured multiplicity distributions, $P(N)$ (for which, as we claim in \cite{JPG}, the corresponding modified combinants $C_j$ oscillate), are most frequently described by Negative Binomial Distributions (the modified combinants of which do not oscillate). Furthermore, with increasing collision energy and increasing multiplicity of produced secondaries some systematic discrepancies between the data and the NBD form of the $P(N)$ used to fit them become more and more apparent. We propose therefore to use a novel phenomenological approach to the observed multiplicity distributions based on the modified combinants  $C_j$ obtained from the measured multiplicity distributions. Together with the fitted multiplicity distributions $P(N)$ they would allow for a more detailed quantitative description of the complex structure of the multiparticle production process. We argue that the observed strong oscillations of the coefficients $C_j$ at the $pp$ data at LHC energies indicate the compound character of the measured distributions $P(N)$ with a central role played by the Binomial Distribution which provides the oscillatory character of the $C_j$. This must be supplemented by some other distribution in such a way that the compound distribution fits both the observed $P(N)$ and $C_j$ deduced from it. However, at the moment we are not able to get fits to both $P(N)$ and $C_j$ of acceptable quality. Therefore, these oscillations still await their physical justification, i.e., identification of some physical process (or combination of such processes) which would result in such phenomenon.

We close by noting that both phenomena discussed here describe, in fact, different dynamical aspects of the multiparticle production process at high energies. The quasi power-like distributions and the related log-periodic oscillations are related with events with rather small multiplicities of secondaries with large and very large momenta; they are called {\it hard collisions} and they essentially probe the collision dynamics towards the edge of the phase space. The multiparticle distributions collect instead all produced particles, the majority of which come from the so called {\it soft collisions} concentrated in the middle of the phase space. In this sense, both the phenomena discussed provide us with complementary new information on these processes and, because of this, they should be considered, as much as possible,jointly\footnote{Because of some similarities observed between  hadronic, nuclear and $e^{+}e^{-}$ collisions \cite{Kittel,Sarkisyan,Bzdak} (see also Chapter $20$ of \cite{PDG2016}), one might expect that the phenomena discussed above will also appear in these reactions. However, this is a separate problem, too extensive and not yet much discussed, to be presented here.}.

\acknowledgments{Acknowledgments}

This research  was supported in part (GW) by the National Science Center (NCN) under contract 2016/22/M/ST2/00176.  We would like to thank warmly Dr Nicholas Keeley for reading the manuscript.

\authorcontributions{Author Contributions}

The content of this article was presented by Z. W\l odarczyk at the SigmaPhi 2017 conference at Corfu, Greece.


\conflictofinterests{Conflicts of Interest}

The authors declare no conflict of interest.

\bibliographystyle{mdpi}

\begin{thebibliography}{999}

\bibitem{Tsallis} Tsallis, C. Possible generalization of Boltzman-Gibbs statistics.
                 {\em  J. Statist. Phys.} {\bf 1998} {\em 52} 479-487.

\bibitem{Tsallis-b} Tsallis, C.  {\em Introduction to Nonextensive Statistical Mechanics} (Springer, 2009).
                    For an updated bibliography on this subject, see http://tsallis.cat.cbpf.br/biblio. htm.

\bibitem{WW_1} Wilk, G.: W{\l}odarczyk, Z.
               Quasi-power law ensembles.
               {\em Acta Phys. Polon. B} {\bf 2015} {\em 46} 1103.

\bibitem{ATLAS-a} Aad, G.; $et~al.$ (ATLAS Collaboration). Charged-particle
                  multiplicities in pp interactions measured with the ATLAS detector at
                  the LHC. {\em New J. Phys.} {\bf 2011}  {\em 13} 053033.

\bibitem{ATLAS-b} Aad, G,; $et~al.$ (ATLAS Collaboration). Properties of jets measured from tracks in proton-proton
                  collisions at center-of-mass energy $\sqrt{s} = 7$ TeV with the ATLAS detector.
                  {\em Phys. Rev. D} {\bf 84} (2011) 054001.

\bibitem{CMS-1} Khachatryan, V.; $et~al.$ (CMS Collaboration). Transverse-momentum and
                pseudorapidity distributions of charged hadrons in pp collisions at
                $\sqrt{s} = 0.9$ and $2.36$ TeV.  {\em J. High Energy Phys.} {\bf 2010} {\em 02} 041.

\bibitem{CMS-2} Khachatryan, V.; $et~al.$ (CMS Collaboration).Transverse-Momentum and
                Pseudorapidity Distributions of Charged Hadrons in pp Collisions
                at $\sqrt{s} =7$ TeV. {\em Phys. Rev. Lett.} {\bf 2010}
                {\em 105} 022002.

\bibitem{WWCT} Wong, C.-T.; Wilk, G.; Cirto, L.J.L.; Tsallis, C.
               From QCD-based hard-scattering to nonextensive statistical mechanical
               descriptions of transverse momentum spectra in high-energy $pp$ and $\bar{p}p$ collisions.
               {\em Phys. Rev. D} {\bf 2015} {\em 91} 114027.

\bibitem{ATLAS-1} Aad, G,; $et~al.$ (ATLAS Collaboration).
                  Measurement of the jet fragmentation function and transverse profile in proton-proton
                  collisions at a center-of-mass energy of $7$ TeV with the ATLAS detector.
                  {\em Eur. Phys. J. C} {\bf 2011} {\em 71} 1795.

\bibitem{AW} Wr\'oblewski, A.
             Multiplicity distributions in proton-proton collisions,
             {\em Acta Phys. Polon. B} {\bf 1973} {\em 4} 857-884.

\bibitem{G-G} Geich-Gimbel, C.
              Particle production at collider energies.
              {\em Int. J. Mod. Phys. A} {\bf 1989} {\em 4} 1527.

\bibitem{ATLAS-2} Aad, G,; $et~al.$ (ATLAS Collaboration).
                  Measurement of the charged particle multiplicity inside jets from
                  $\sqrt{s} = 8$ TeV $pp$ collisions with the ATLAS detector.
                  {\em Eur. Phys. J. C} {\bf 2016} {\em 76} 322.

\bibitem{WW_a} Wilk, G.; W{\l}odarczyk, Z.
               Self-similarity in jet events following from pp collisions at LHC.
               {\em Phys. Lett. B} {\bf 2013} {\em 727} (2013) 163-167.

\bibitem{WW_b} Wilk, G.; W{\l}odarczyk, Z.
               Power laws in elementary and heavy-ion collisions.
               {\em Eur. Phys. J. A} {\bf 2009} {\em 40} (2013) 299-312.

\bibitem{Hagedorn} Hagedorn, R.; Ranft, R.
                   Statistical thermodynamics of strong interactions at high energies. II - Momentum spectra
                   of particles produced in $pp$ collisions.
                   {\em Suppl. Nuovo Cim.} {\bf 1968} {\em 6} 169-310.

\bibitem{Hagedorn-a} Hagedorn, R.
                     Remarks of the thermodynamical model of strong interactions.
                     {\em Nucl. Phys. B} {\bf 1970} {\em 24} 93-139.

\bibitem{AD} Deppman, A.
             Thermodynamics with fractal structure, Tsallis statistics, and hadrons.
             {\em Phys. Rev. D} {\bf 2016} {\em 93} 054001.

\bibitem{SSofNBD} Calucci, G; Treleani, D.
                  Self-similarity of the negative binomial multiplicity distributions.
                  {\em Phys. Rev. D} {\bf 1998} {\em 57} 602-605.

\bibitem{DeWolf} De Wolf, E. A.; Dremin, I. M.; Kittel, W.
                 Scaling laws for density correlations and fluctuations and fluctuations in multiparticle dynamics.
                 {\em Phys. Rep.} {\bf 1996} {\em  270} 1.

\bibitem{Kittel} Kittel, W.; De Wolf, E. A.;
                 Soft Multihadron Dynamics,
                 World Scientific, Singapore {\bf 2005}.

\bibitem{CMS-3} Chatrchyan, S.; $et~al.$ (CMS Collaboration).
                Charged particle transverse momentum spectra in $pp$ collisions at $\sqrt{s} = 0.9$ and $7$ TeV.
                {\em J. High Energy Phys.} {\bf 2011} {\em 08} 086.

\bibitem{ATLAS-3} Aad, G,; $et~al.$ (ATLAS Collaboration).
                  Charged-particle multiplicities
                  in $pp$ interactions measured with the ATLAS detector at the LHC.
                  {\em New J. Phys.} {\bf 2011} {\em 3} 053033.

\bibitem{WW_c} Wilk, G.; W{\l}odarczyk, Z.
               Temperature oscillations and sound waves in hadronic matter.
               {\em Physica A} {\bf 2017} {\em 486} 579-586.

\bibitem{Scaling} Sornette, D.
                  Discrete-scale invariance and complex dimensions.
                  {\em Phys. Rep.} {\bf 1998} {\em 297}  239-270.

\bibitem{WW_2} Wilk, G.; W{\l}odarczyk, Z.
               Tsallis distribution with complex nonextensivity parameter $q$.
               {\em Physica A} {\bf 2014} {\em 413}, 53-58.

\bibitem{RWW} Rybczy\'nski, M.; Wilk, G.; W{\l}odarczyk, Z.
              System size dependence of the log-periodic oscillations of transverse momentum spectra.
              {\em EPJ Web of Conf.} {\bf 2015} {\em 90} 01002.

\bibitem{WW_3} Wilk, G.; W{\l}odarczyk, Z.
               Tsallis Distribution Decorated with Log-Periodic Oscillation.
               {\em Entropy} {\bf 2015} {\em 17} 384-400.

\bibitem{WW_4} Wilk, G.; W{\l}odarczyk, Z.
               Quasi-power laws in multiparticle production processes.
               {\em Chaos Solit. Frac.} {\bf 2015} {\em 81} 487-496.

\bibitem{WW_d} Wilk, G.; W{\l}odarczyk, Z.
               Nonextensive information entropy for stochastic networks.
               {\em Acta Phys. Polon. B} {\bf 2004} {\em 35} 871-879.

\bibitem{Landau} Landau, L.D.; Lifshitz, E.M.
                 Fluid mechanics, Pergamon Press, Oxford {\bf 1987} .

\bibitem{BB} Barenblatt, G.I.
             Scaling, Self-Similarity, and Intermediate Asymptotics,
             Cambridge University Press, {\bf 1996}.

\bibitem{JPG} Wilk, G.; W{\l}odarczyk, Z.
              How to retrieve additional information from the multiplicity distributions.
              {\em J. Phys. G} {\bf 2017} {\em 44} 015002.

\bibitem{GU} Giovannini, A.; Ugocciono, R.
             Signals of new physics in global event properties in pp collisions in the TeV energy domain.
             {\em Phys. Rev. D} {\bf 2003} {\em 68} 034009.

\bibitem{Z} Zborovsky, I. J.
            A three-component description of multiplicity distributions in pp collisions at the LHC.
            {\em J. Phys. G} {\bf 2013} {\em 40} 055005.

\bibitem{DN} Dremin, I. M.; Nechitailo, V. A.
             Independent pair parton interactions model of hadron interactions.
             {\em Phys. Rev. D} {\bf 2004} {\em 70} 034005.

\bibitem{DG} Dremin, I. M.; Gary, J. W.
             Hadron multiplicities.
             {\em Phys. Rep.} {\bf 2001} {\em 349} 301.

\bibitem{G-OR} Grosse-Oetringhaus, J. F.; Teygers, K.
               Charged-particle multiplicity in proton–proton collisions.
               {\em J. Phys. G} {\bf 2010} {\em 37} 083001.

\bibitem{CSP} Saleh, B. E, A.; Teich, M. K.
              Multiplied-Poisson Noise in Pulse, Particle, and Photon Detection.
              {\em Proc. IEEE}  {\bf 1982} {\em 70} 229-245.

\bibitem{CMS-4} V. Khachatryan, V; $et~al.$ (CMS Collaboration).
                Charged particle multiplicities in $pp$ interactions at $\sqrt{s} = 0.9,~2.36$, and $7$ TeV.
                {\em J. High Energy Phys.} 01 {\bf 2011} 79.

\bibitem{NBD-PG} Ghosh, P.
                 Negative binomial multiplicity distribution in proton-proton collisions in
                 limited pseudorapidity intervals at LHC up to $\sqrt{s} = 7$ TeV and the clan model.
                 {\em Phys. Rev. D} 85 {\bf 2012} 054017.

\bibitem{KG}  Kauffmann, S. K.; Gyulassy, M.
              Multiplicity distributions of created bosons: the method of combinants.
             {\em J. Phys. A} {\bf 1978} {\em 11} 1715-1727.

\bibitem{CombUse} Balantekin, A. B.; J.E. Seger, J. E.
                  Description of pion multiplicities using combinants.
                  {\em Phys. Lett. B} {\bf 1991} {\em 266} 231-235.

\bibitem{H} Hegyi, S.
            Correlation studies in quark jets using combinants.
            {\em Phys. Lett. B} {\bf 1999} {\em 463} 126-131.

\bibitem{BSWW} Biyajima, M.; Suzuki, N.; Wilk, G; W\l odarczyk, Z.
               Totally chaotic poissonian-like sources in multiparticle production processes?
               {\em Phys. Lett. B} {\bf 1996} {\em 386} 297-303.

\bibitem{GVH} Giovannini, A.; Van Hove, L.
              Negative Binomial Multiplicity Distributions in High Energy Hadron Collisions.
              {\em Z. Phys. C} {\bf 1986} {\em 30} 391.

\bibitem{Compound} Sundt, B.; Vernic, R.
                   Recursions for Convolutions and Compound Distributions with Insurance Applications,
                   Springer-Verlag Berlin Heidelberg {\bf 2009}.

\bibitem{Panjer} Panjer, H.H.
                 Recursive evaluation of a family of compound distributions.
                 {\em ASTIN Bull.} {\bf 1981} {\em 12} 22-26.

\bibitem{PNAS} Flindt, C.; Fricke, C.; Hohls, F.; Novotny, T.; Netocny, K.; Brandes, T.; Haug, R.J.
               Universal oscillations in counting statistics.
               {\em Proc. Natl. Acad. Sci. U. S. A.} {\bf 2009} {\em 106} 10116-10119.

\bibitem{Sarkisyan} Sarkisyan, E. K.; Mishra, A. N.; Sahoo, R.; Alexander S. Sakharov, A. S.
                    Multihadron production dynamics exploring the energy balance in hadronic and nuclear collisions.
                    {\em Phys. Rev. D} {\bf 2016} {\em 93} 054046.

\bibitem{Bzdak} Bzdak, A.
                Universality of multiplicity distribution in proton-proton and electron-positron collisions
                {\em Phys. Rev. D} {\bf 2017} {\em 96} 036007.

\bibitem{PDG2016} Patrignani, C. $et~al.$  (Particle Data Group)
                  Review of particle physics.
                  {\em Chinese Phys. C} {\bf 2016} {\em 40} 100001.


\end{thebibliography}
\makeatletter
\renewcommand\@biblabel[1]{#1. }
\makeatother

\end{document}